\documentclass[a4paper,fleqn,sort&compress]{cas-dc}
\usepackage[numbers]{natbib}
\usepackage{amsmath}
\usepackage{amssymb}
\usepackage{graphics}
\usepackage{epstopdf}
\usepackage{hyperref}

\def\tsc#1{\csdef{#1}{\textsc{\lowercase{#1}}\xspace}}
\tsc{WGM}
\tsc{QE}

\begin{document}
\let\WriteBookmarks\relax
\def\floatpagepagefraction{1}
\def\textpagefraction{.001}

\shorttitle{}    

\shortauthors{D. Siv\'{y} and J. Stre\v{c}ka}  

\title [mode = title]{Field-driven quantum phase transitions in a spin-$1/2$
Heisenberg antiferromagnet on an extended Lieb lattice}  



%

\author[1]{D. Siv\'{y}}[orcid=0000-0001-9486-7470]



\credit{Writing – original draft, Methodology, Investigation,
Formal analysis, Data curation}

\affiliation[1]{organization={Institute of Physics, Faculty of Science, P. J. \v{S}af\'{a}rik University},
            addressline={Park Angelinum 9}, 
            postcode={040 01}, 
            postcodesep={},
            city={Ko\v{s}ice},
            country={Slovakia}}

\author[1]{J. Stre\v{c}ka}[orcid=0000-0003-1667-6841]

\cormark[1]

\ead{jozef.strecka@upjs.sk}

\credit{Writing – review \& editing,  Methodology, Investigation, Formal analysis, Data
curation, Validation, Supervision, Funding acquisition, Conceptualization}

\cortext[1]{Corresponding author}



\begin{abstract}
The magnetic behavior of the spin-1/2 Heisenberg antiferromagnet on the extended Lieb lattice is examined in  the presence of an external magnetic field. Using density matrix renormalization group calculations, we determine the zero-temperature magnetization curves and construct the ground-state phase diagram in the magnetic field vs. the interaction ratio plane. Depending on the interaction ratio, the zero-temperature magnetization curves exhibit field-driven quantum phase transitions between the gapped phases manifested as the intermediate plateaus at $1/5$ and $3/5$ of the saturation magnetization and a gapless spin-canted phase, in which the magnetization varies continuously with the applied magnetic field. Quantum Monte Carlo simulations reveal clear finite-temperatures signatures of the field-induced quantum phase transitions and the intermediate magnetization plateaus, while they provide no evidence for either continuous or discontinuous thermal phase transitions.
\end{abstract}


\begin{highlights}
\item Ground-state phase diagram of the Heisenberg model on the extended Lieb lattice.
\item Field-induced quantum phase transitions generate 1/5 and 3/5 magnetization plateaus.
\item Gapless spin-canted phase separates gapped plateau phases.
\item QMC reveals thermal signatures of quantum criticality but no thermal transitions.
\end{highlights}


\begin{keywords}
extended Lieb lattice \sep quantum phase transition \sep Heisenberg model \sep quantum Monte Carlo \sep magnetization plateau
\end{keywords}

\maketitle

\section{Introduction}\label{intro}

Competing interactions that cannot be simultaneously satisfied in two-dimensional (2D) spin systems give rise to geometric spin frustration \cite{{lieb86},{die04}} leading to a variety of nontrivial phenomena as exemplified by the prototypical frustrated 2D Ising model \cite{{lieb86},{die04},{kal11},{jin12}}. The Heisenberg model on a diamond-decorated square lattice represents a paradigmatic example of frustrated 2D quantum spin system, which exhibits in an external magnetic field a line of discontinuous thermal phase transitions terminating at an Ising critical point  \cite{cac23}. This remarkable behavior is thus particularly intriguing from the perspective of phase transitions and critical phenomena. More generally, the introduction of geometric frustration into quantum 2D spin models subjected to an external magnetic field has been shown to significantly enrich their phase transitions \cite{{sen04},{lac10},{fan24}}.

This naturally raises the question of whether a non-frustrated 2D quantum spin model in an external magnetic field can exhibit similar thermal phase transitions. In our recent study, we investigated a non-frustrated spin-1/2 Ising-Heisenberg model on the extended Lieb lattice subjected to an external magnetic field \cite{siv26}. By utilizing the decoration-iteration transformation \cite{fis59}, this model was mapped exactly onto an effective classical Ising model on a square lattice revealing a rich variety of nontrivial thermal phase transitions. A key open question is whether lifting the classical constraints and introducing full quantum fluctuations destroys these thermal phase transitions. Motivated by this question, the present work focuses on the fully quantum spin-1/2 Heisenberg antiferromagnet on the same extended Lieb lattice in the presence of an external magnetic field.

\section{Model and methods}\label{model}

\begin{figure}
  \centering
    \includegraphics[scale=0.08]{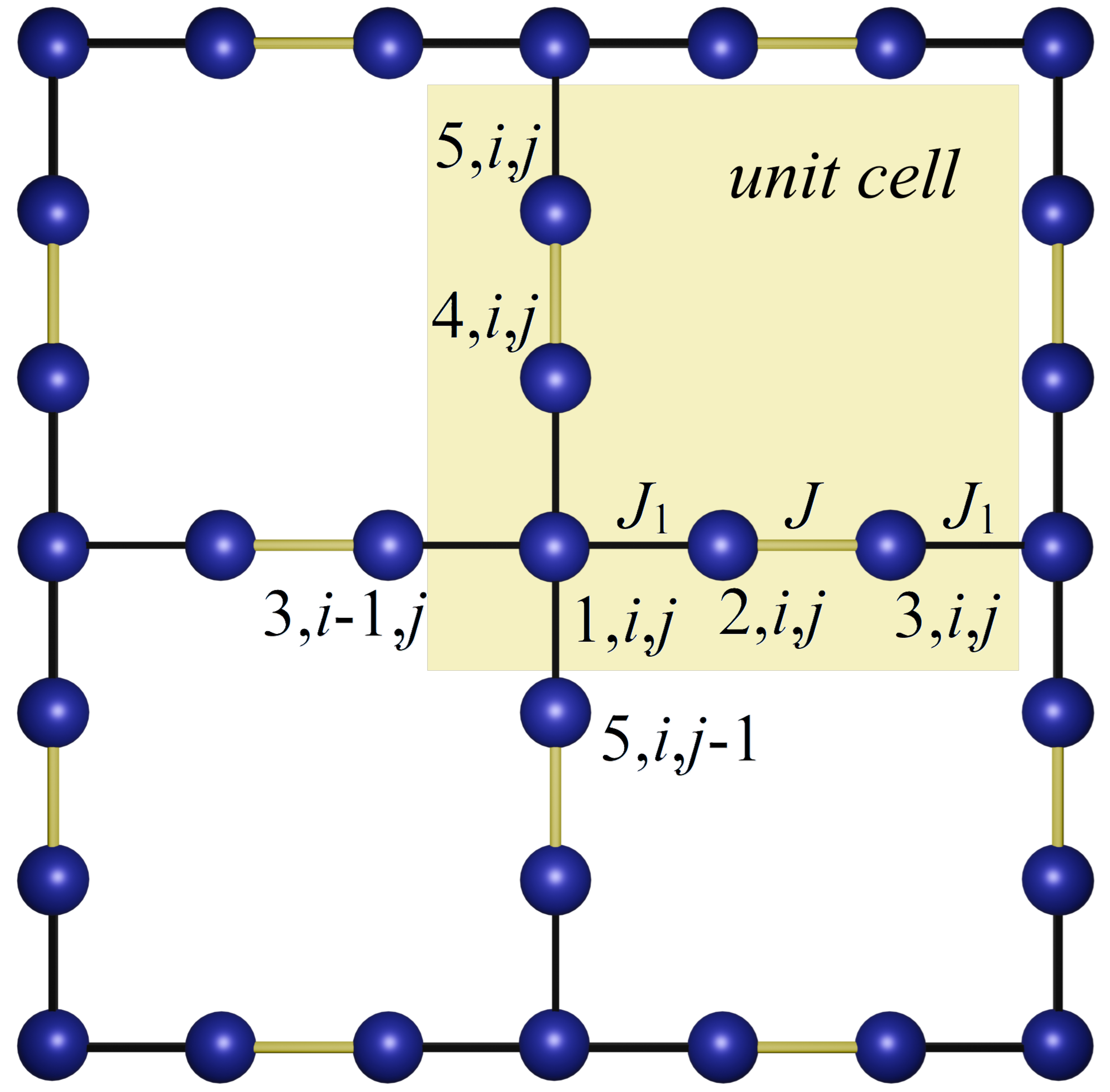}
    \caption{Schematic illustration of the spin-1/2 Heisenberg model on the extended Lieb lattice, where blue circles represent the lattice sites occupied by Heisenberg spins. The five
spins constituting the unit cell are also indicated.}\label{fig1}
\end{figure}

We consider the spin-$1/2$ Heisenberg antiferromagnet on the extended Lieb lattice in a magnetic field with periodic boundary conditions. The Hamiltonian of the model reads
\begin{align}
\hat{\mathcal{H}} &= 
 J_1\! \sum_{i,j=1}^{L}\left[
 \boldsymbol{\hat{S}}_{1,i,j} \cdot \left( \boldsymbol{\hat{S}}_{2,i,j} + \boldsymbol{\hat{S}}_{3,i-1,j} + \boldsymbol{\hat{S}}_{4,i,j} + \boldsymbol{\hat{S}}_{5,i,j-1} \right)\right]  \nonumber \\
&+ J \!\!\sum_{i,j=1}^{L} \!\left( \!\boldsymbol{\hat{S}}_{2,i,j} \!\cdot\! \boldsymbol{\hat{S}}_{3,i,j} \!+\! \boldsymbol{\hat{S}}_{4,i,j} \cdot \boldsymbol{\hat{S}}_{5,i,j} \!\right) \!-\! h\!\! \sum_{i,j=1}^{L} \sum_{k=1}^{5}\! \hat{S}_{k,i,j}^{z},
\label{ham}
\end{align}
where $\boldsymbol{\hat{S}}_{k,i,j}\equiv (\hat{S}_{k,i,j}^x,\hat{S}_{k,i,j}^y,\hat{S}_{k,i,j}^z)$ are the spin-$1/2$ operators representing the $k$th spin in the $(i,j)$th unit cell, $L$ is the linear lattice size, and the coupling constants $J$ and $J_1$ represent two different exchange interactions schematically illustrated in Fig. \ref{fig1}. The Zeeman coupling of the spins to the external magnetic field $h$ is represented by the last term of Hamiltonian (\ref{ham}). Note that the indices $i-1,j$ ($i,j-1$) correspond to the unit cell to the left (below) with respect to the $(i,j)$th unit cell.

To investigate the spin-$1/2$ Heisenberg antiferromagnet on the extended Lieb lattice given by the Hamiltonian (\ref{ham}), we employ density matrix renormalization group (DMRG) calculations and stochastic series expansion quantum Monte Carlo (QMC) simulations, both implemented within the Algorithms and Libraries for Physics Simulations (ALPS) project \cite{bau11}. The DMRG method is used to obtain zero-temperature magnetization curves and to construct the ground-state phase diagram. Within the DMRG calculations, we retained up to 2000 kept states and performed up to 20 sweeps for systems with linear sizes up to $L=6$ (180 spins). On the other hand, QMC simulations allows us to study typical features of the magnetization and the magnetic susceptibility at finite temperatures. Within QMC simulations, we used up to $5 \times 10^5$ Monte Carlo steps for systems with linear sizes up to $L=8$ (320 spins) with the first 20 \%  of the steps discarded for thermalization and the remaining 80 \% used for statistical averaging. Throughout this work,  the size of the coupling constant $J$ will serve as the unit of energy. 

\section{Results and discussion}\label{results}

\begin{figure*}
\centering
\includegraphics[scale=0.28]{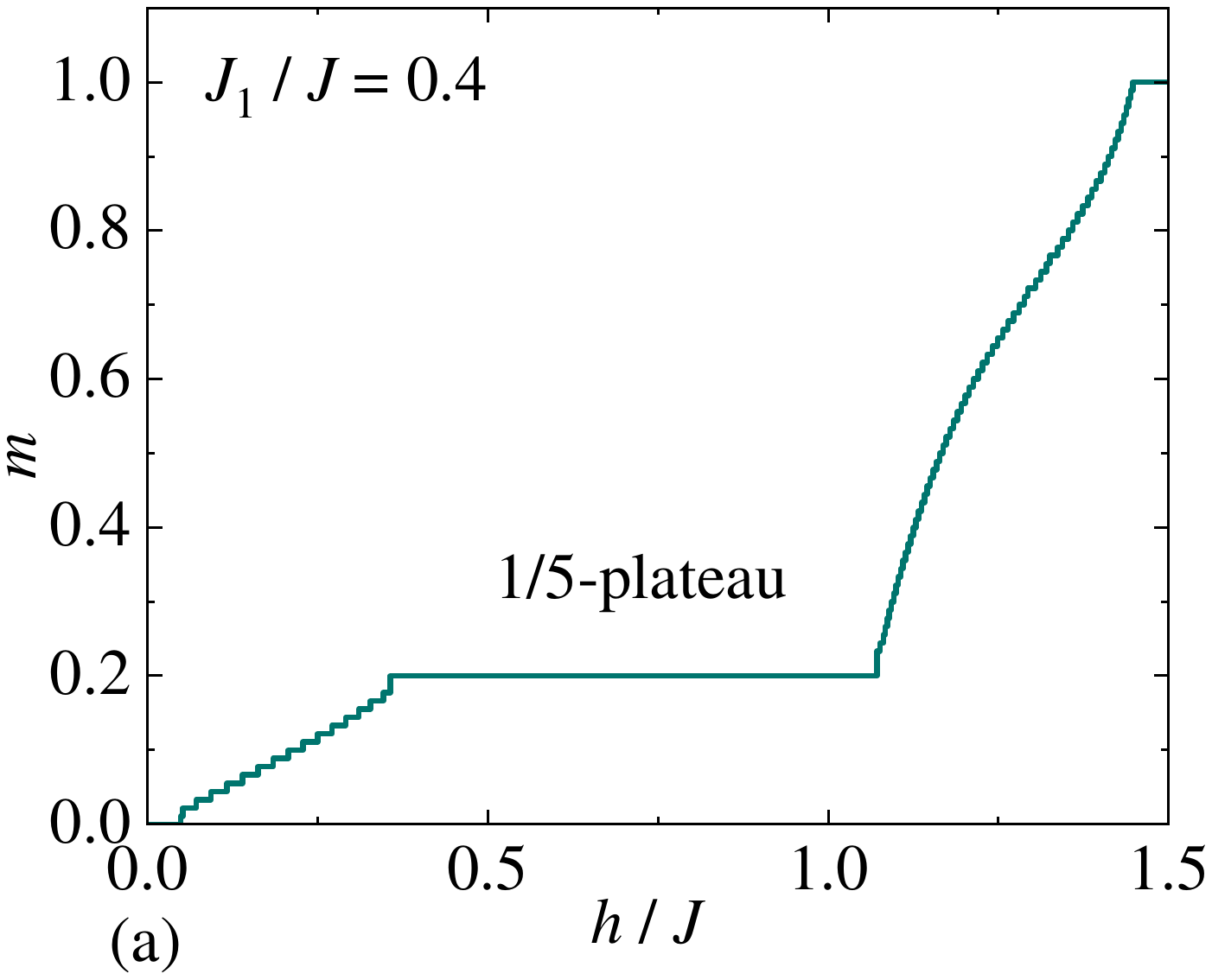}
\includegraphics[scale=0.28]{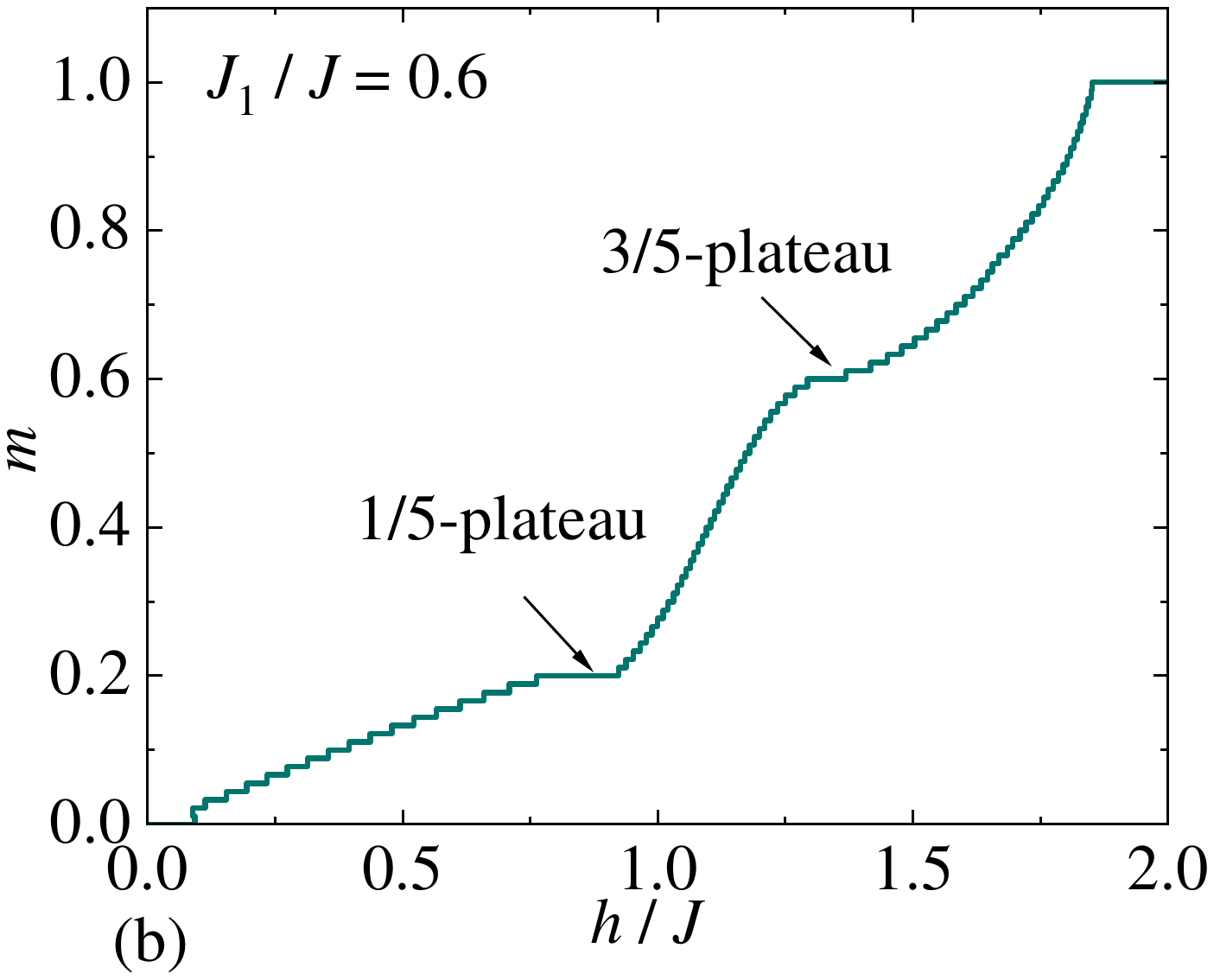}\\[-0.5cm]
\includegraphics[scale=0.28]{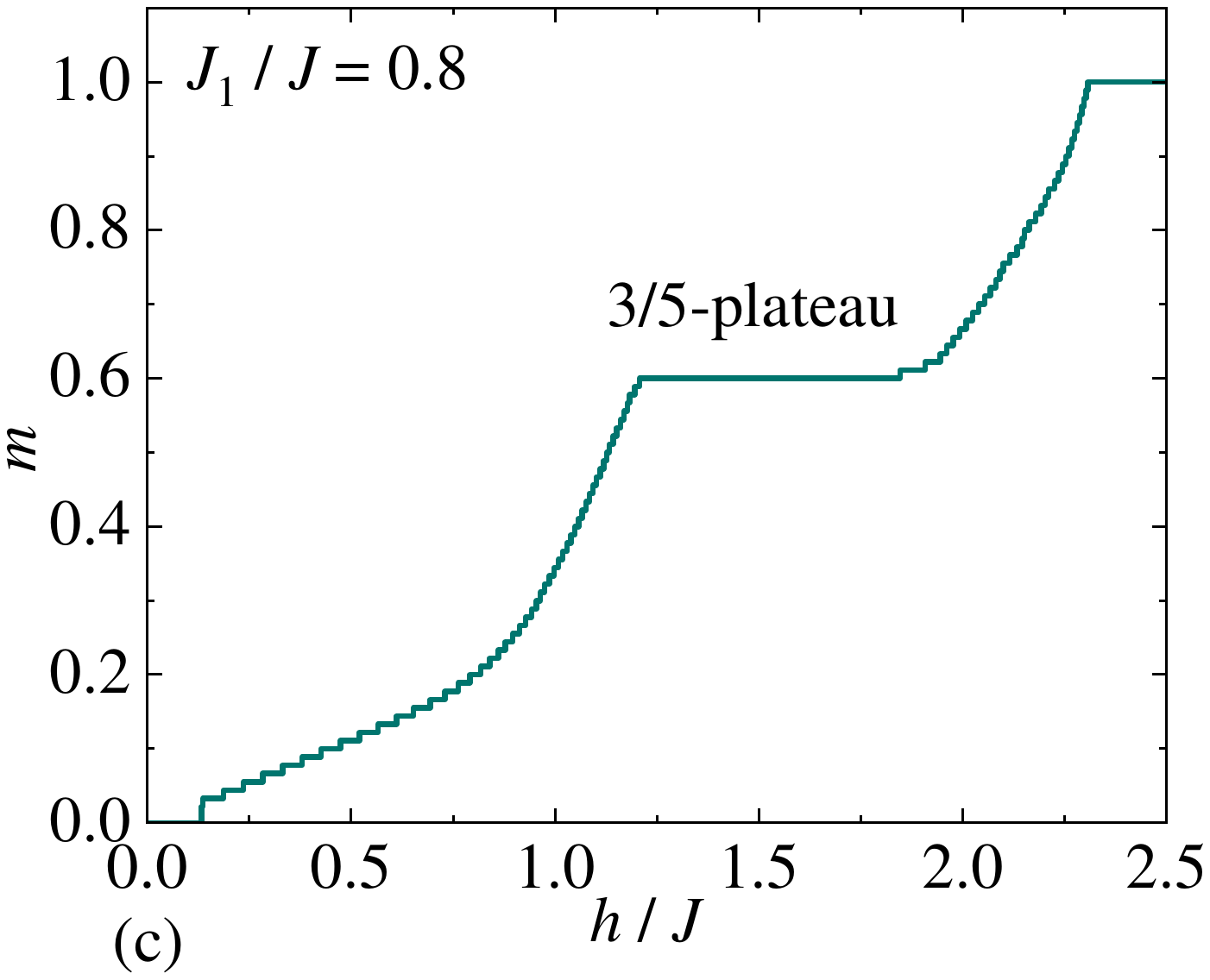}
\includegraphics[scale=0.28]{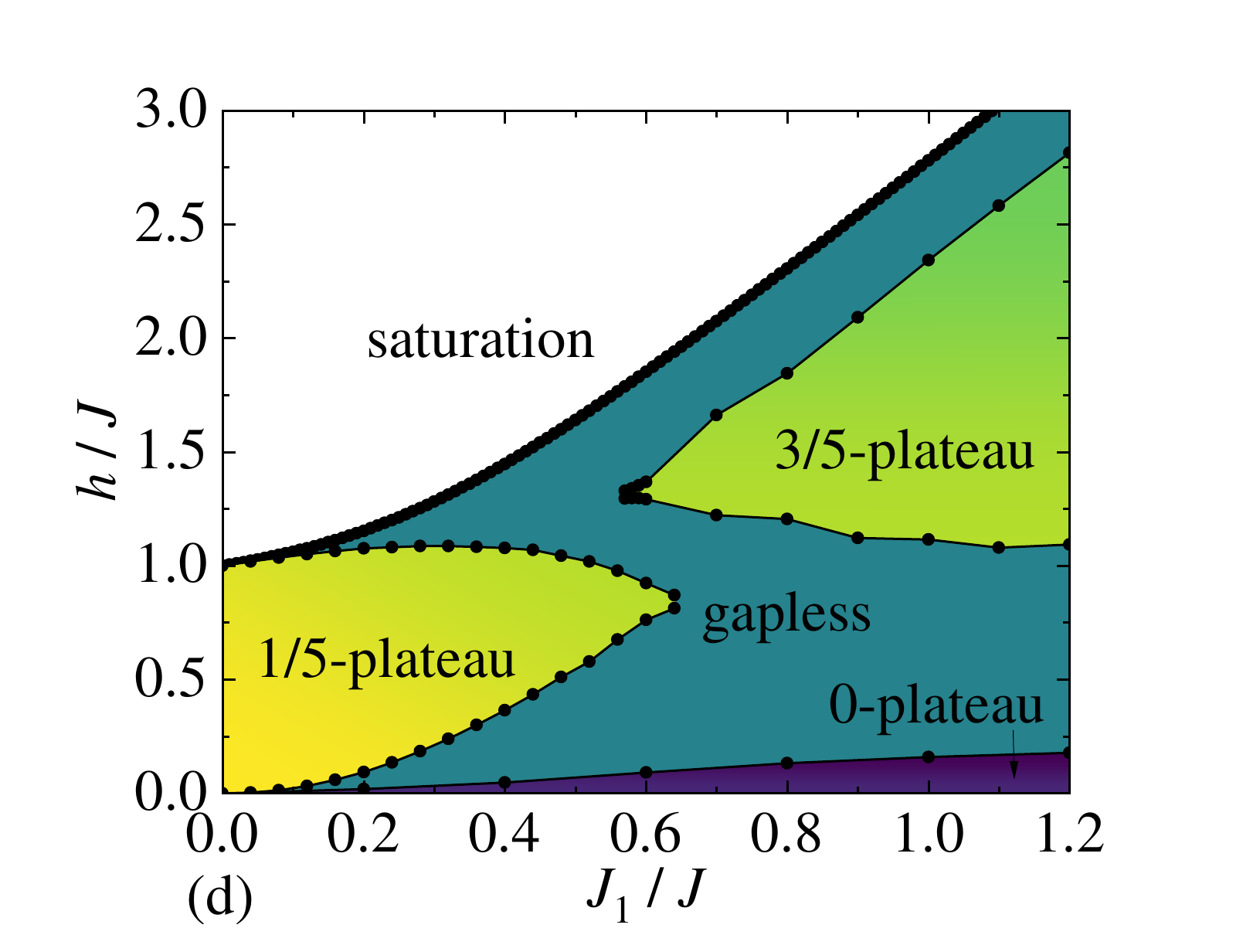}
\caption{Field dependencies of the magnetization of the spin-1/2 Heisenberg on the extended Lieb lattice at zero temperature	 and three selected values of the interaction ratio: (a) $J_1/J = 0.4$; (b) $J_1/J = 0.6$; (c) $J_1/J = 0.8$. (d) The ground-state phase diagram in the $J_1/J-h/J$ plane with the phases indicated. The results were obtained using DMRG simulations for a linear system size of $L = 6$ corresponding to 180 spins.}
\label{fig2}
\end{figure*}

In this section, we present the key findings for the spin-$1/2$ Heisenberg antiferromagnet on the extended Lieb lattice. Fig.~\ref{fig2}(a)–(c) illustrate three typical zero-temperature magnetization curves. For the lowest interaction ratio $J_1/J = 0.4$, the magnetization exhibits a narrow zero-magnetization plateau followed by a multistep increase indicating a gapless spin-canted regime. In the thermodynamic limit,  this multistep behavior is expected to evolve into a smooth continuous increase of the magnetization within the spin-canted phase until the system undergoes a field-induced quantum phase transition (QPT) into a gapped phase, which manifest itself through an intermediate $1/5$ plateau starting nearly at $h/J \approx 0.36$ [see Fig.~\ref{fig2}(a)]. The gapful 1/5-plateau phase persists up to a second field-induced QPT emerging at the upper critical field $h/J \approx 1.07$, beyond which the system re-enters the gapless spin-canted regime repeatedly reflected in a multistep dependence before the magnetization eventually reaches the full saturation. The zero-temperature magnetization curve shown in Fig.~\ref{fig2}(b) for $J_1/J = 0.6$ reveals the emergence of an additional intermediate $3/5$ plateau approximately at $h/J \approx 1.3$, while the width of the intermediate $1/5$ plateau is substantially reduced. For even higher values of the interaction ratio such as $J_1/J = 0.8$, the $1/5$ plateau disappears completely from the zero-temperature magnetization curve, whereas the $3/5$ plateau becomes significantly more pronounced as illustrated in  Fig.~\ref{fig2}(c).

Next, we analyze the ground-state phase diagram shown in Fig.~\ref{fig2}(d) in the $h/J$–$J_1/J$ plane, which is constructed from the zero-temperature magnetization curves of the spin-$1/2$ Heisenberg antiferromagnet on the extended Lieb lattice with the linear system size $L=6$ (i.e. 180 spins). The diagram comprises three gapped phases corresponding to the $0$, $1/5$, and $3/5$ magnetization plateaus, a gapless spin-canted phase, and the fully saturated phase. As the interaction ratio increases, the intermediate $1/5$ plateau progressively reduces until it is fully suppressed by the gapless spin-canted phase, whereas the intermediate $3/5$ plateau contrarily develops and successively broadens with increasing the interaction ratio. It should be emphasized that all results presented here were obtained for a finite system with linear size $L=6$ corresponding to 180 spins. Whether or not the narrow zero-magnetization plateau survives in the thermodynamic limit thus remains an open question. Resolving this issue would require a systematic finite-size scaling analysis, which is beyond the scope of the present work.  On the other hand, the existence of the considerably more robust intermediate $1/5$ and $3/5$ magnetization plateaus is evident even from the results got the present system size.  A more extensive finite-size analysis would primarily refine only the precise locations of the corresponding plateau regions including the interaction ratio at which the $1/5$ plateau vanishes and the $3/5$ plateau first appears. Unlike the spin-1/2 Heisenberg diamond-decorated square lattice representing frustrated counterpart of the present model, the present unfrustrated model provides no evidence of a direct phase boundary separating the gapped 1/5- and 3/5-plateau phases. Consequently, the finite-temperature phase transition associated with such a phase boundary in the frustrated model is likely absent in the present unfrustrated model. 

\begin{figure}
\centering
\includegraphics[scale=0.28]{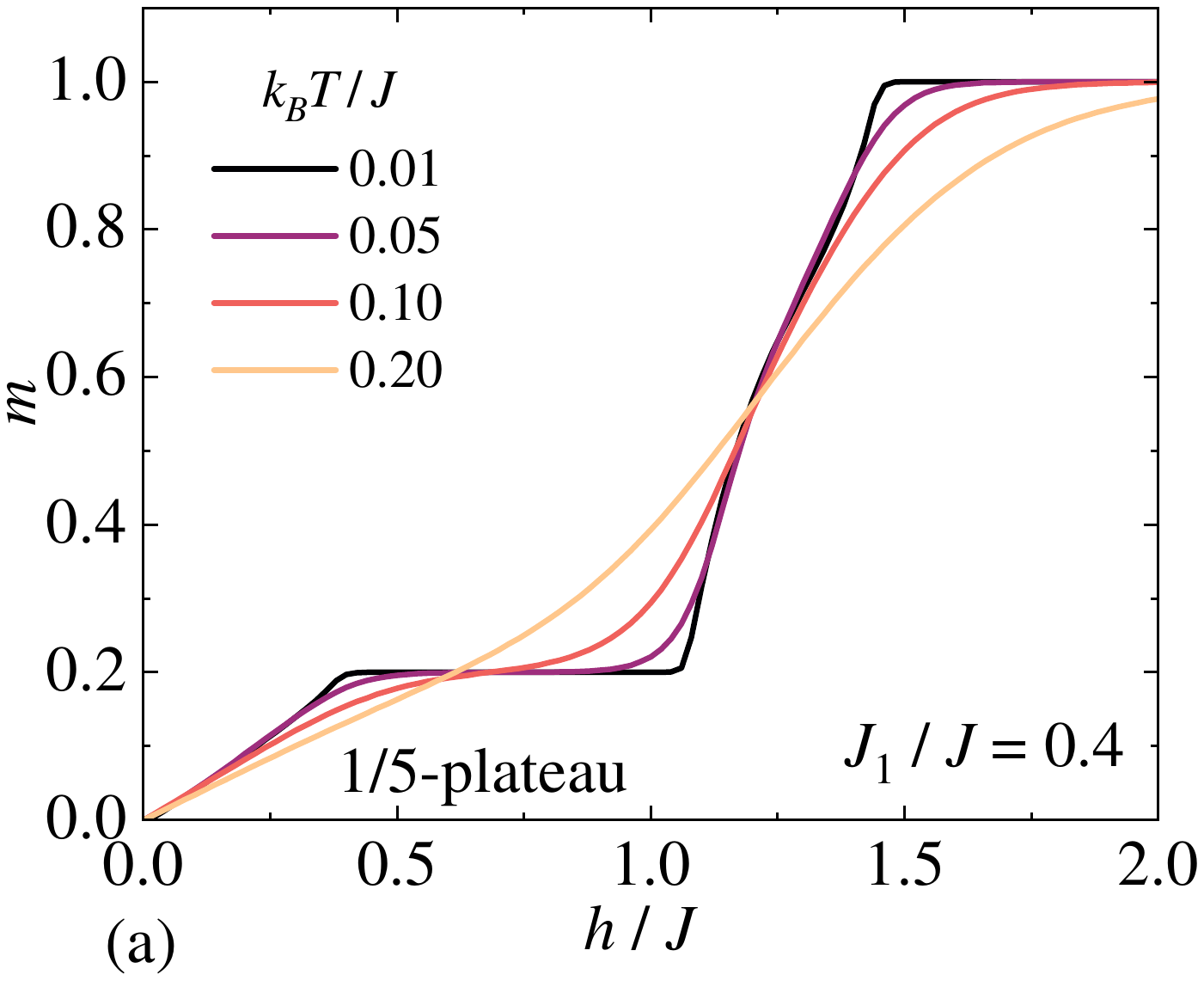}\\[-0.5cm]
\includegraphics[scale=0.28]{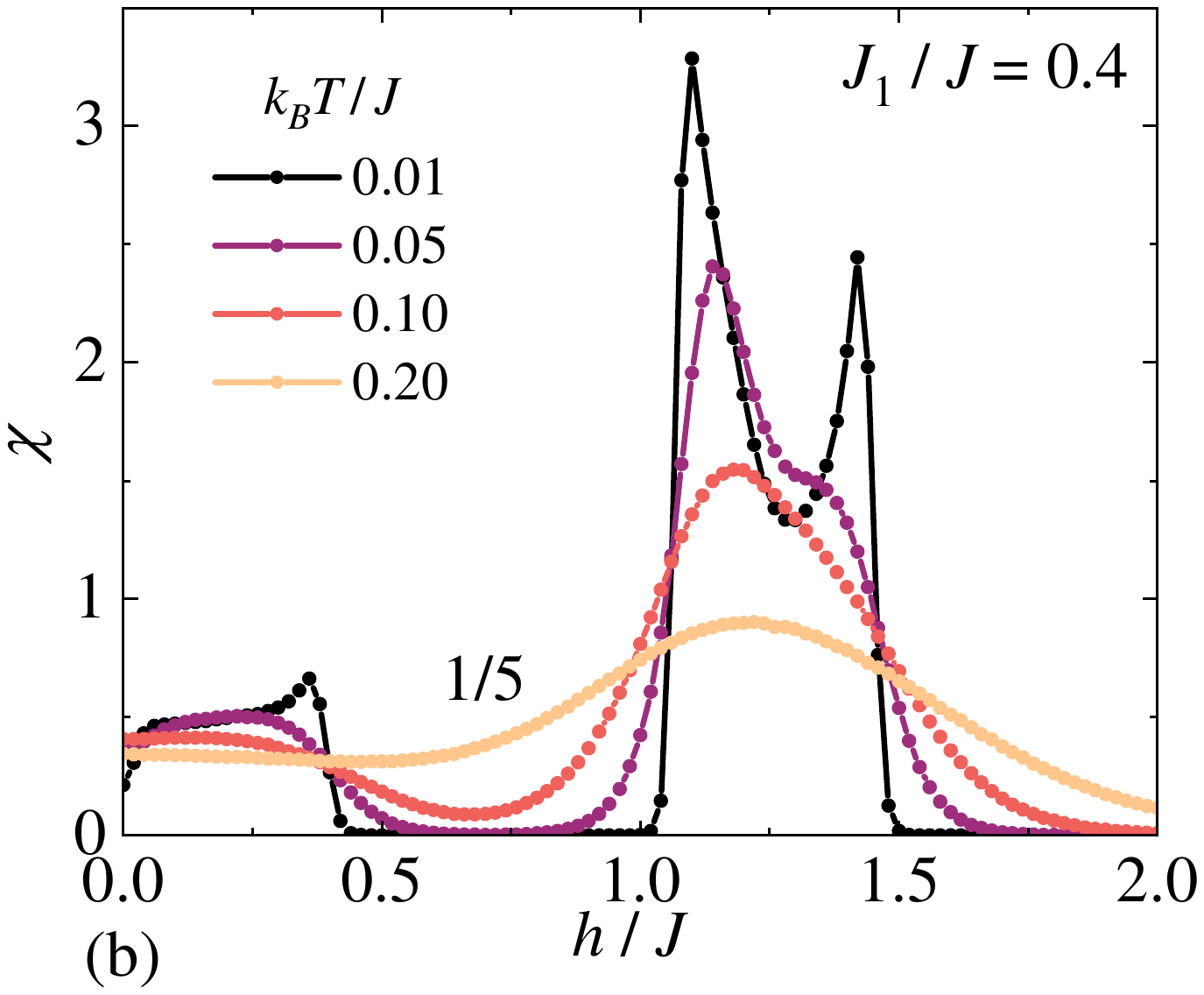}
\caption{Field dependencies of the magnetization normalized with respect to its saturation value (a) and the corresponding magnetic susceptibility (b) for the spin-1/2 Heisenberg model on the extended Lieb lattice at four distinct temperatures for $J_1/J=0.4$ as obtained from QMC simulations for the linear system size $L=8$ (320 spins).}
\label{fig4}
\end{figure}

To further examine this issue, we investigated the finite-temperature magnetic properties of the spin-$1/2$ Heisenberg model on the extended Lieb lattice using QMC simulations for a larger system with linear size $L=8$ (320 spins). Fig.~\ref{fig4}(a) and \ref{fig4}(b) show the field dependence of the magnetization and magnetic susceptibility at four different temperatures for $J_1/J = 0.4$. The isothermal magnetization curve at the lowest considered temperature $k_{\rm B}T/J = 0.01$ is fully consistent with the zero-temperature magnetization data obtained from DMRG simulations including a sizable intermediate 1/5 plateau ranging in between the lower and upper critical fields $h/J \approx 0.4$ and $h/J \approx 1.1$, respectively. A gradual melting of the magnetization curve is then observed upon increasing temperature, whereby the intermediate $1/5$ plateau is fully suppressed and the magnetization approaches its saturation value smoothly without any discernible anomalies at sufficiently high temperatures (e.g., $k_{\rm B}T/J = 0.2$). 

The magnetic susceptibility reveals even more subtle anomalies as demonstrated in Fig.~\ref{fig4}(b). At the lowest temperature $k_{\rm B}T/J = 0.01$, the susceptibility displays a small peak near $h/J \approx 0.36$ and a second considerably more pronounced peak near $h/J \approx 1.1$. The susceptibility suddenly drops nearly to zero between these two peaks signaling the existence of the gapped $1/5$-plateau phase. A third susceptibility peak near $h/J \approx 1.4$ together with the two preceding ones can be interpreted as residual signatures of the three field-induced QPTs, because they all are substantially reduced in magnitude with increasing temperature. The first peak at $h/J \approx 0.36$ becomes barely distinguishable already at temperature $k_{\rm B}T/J = 0.05$, while the third peak evolves into a shoulder superimposed on the most pronounced second peak. At sufficiently high temperatures (e.g., $k_{\rm B}T/J = 0.2$), the susceptibility profile becomes smooth and exhibits only a single broad maximum consistent with the continuous rise of the magnetization. It should be noted that the apparent sharpness of the low-temperature peaks in the susceptibility is partly a consequence of the finite field step used in the QMC simulations.

\begin{figure}
\centering
\includegraphics[scale=0.28]{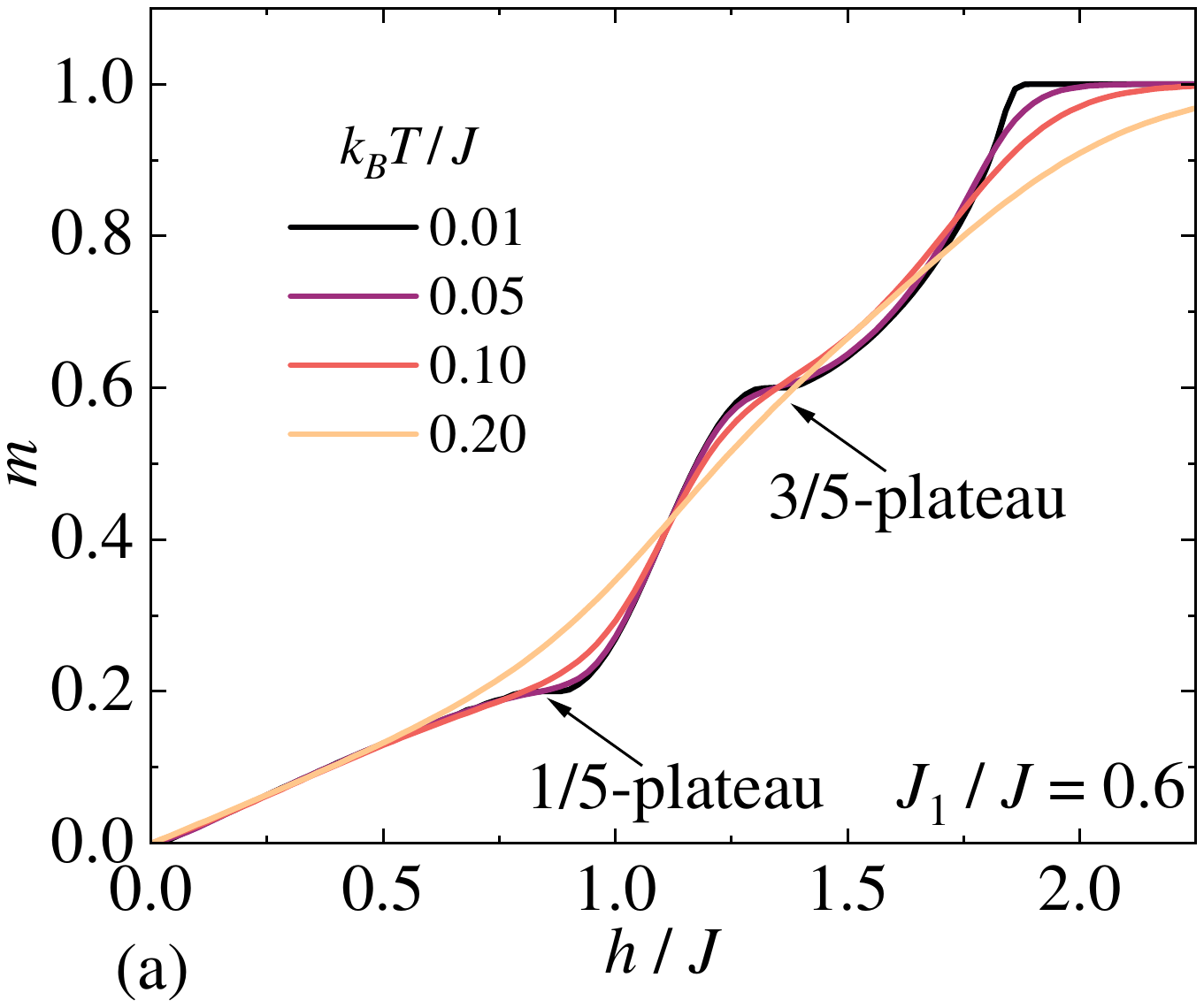}\\[-0.5cm]
\includegraphics[scale=0.28]{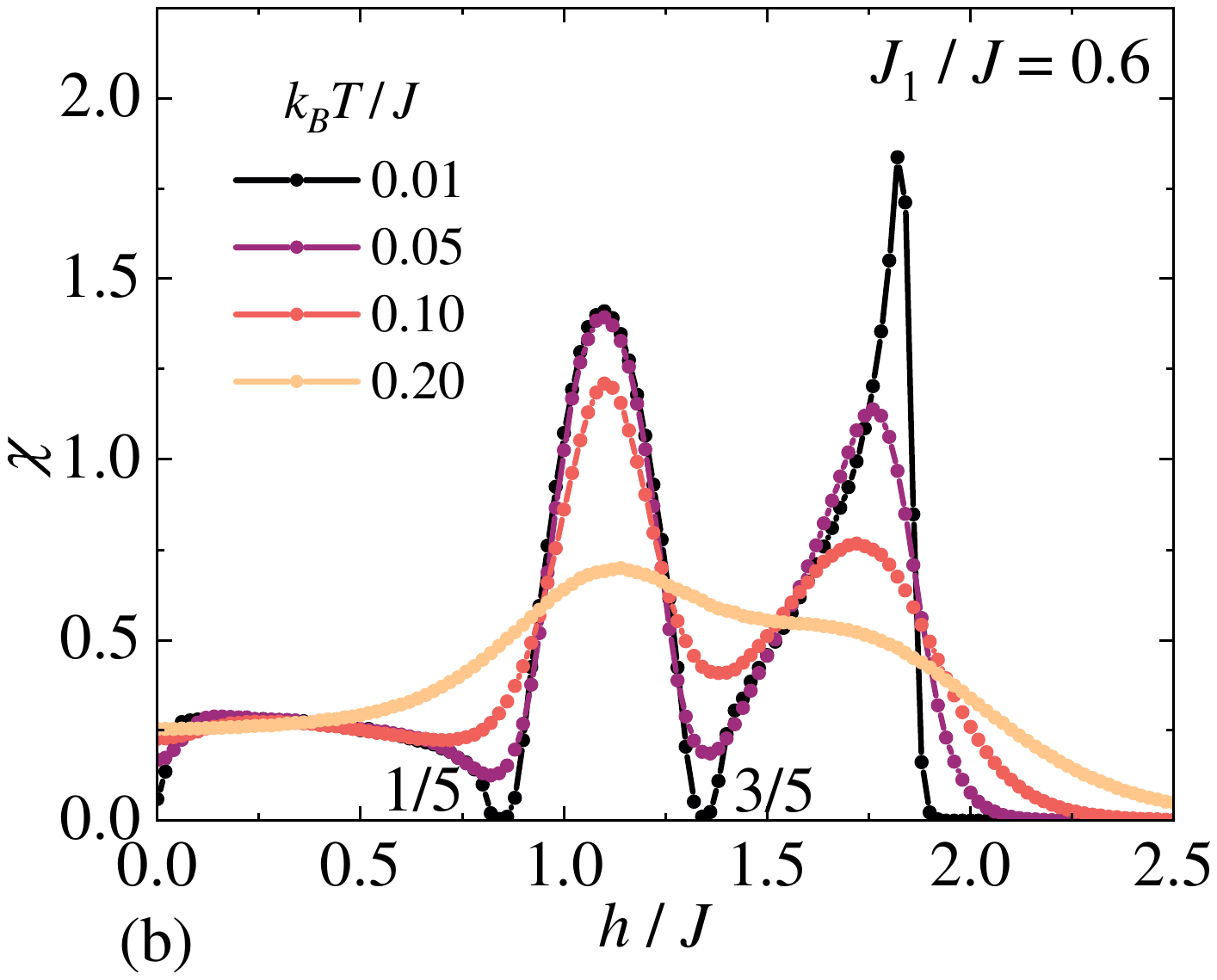}
\caption{Field dependencies of the magnetization normalized with respect to its saturation value (a) and the corresponding magnetic susceptibility (b) for the spin-1/2 Heisenberg model on the extended Lieb lattice at four distinct temperatures for $J_1/J=0.6$ as obtained from QMC simulations for the linear system size $L=8$ (320 spins).}
\label{fig5}
\end{figure}

Next, we consider the case of $J_1/J=0.6$, for which the magnetization and magnetic susceptibility are plotted in Fig.~\ref{fig5} against the magnetic field at four different temperatures. The isothermal magnetization curve depicted in Fig.~\ref{fig5}(a) at the lowest temperature $k_{\rm B}T/J = 0.01$ exhibits two relatively narrow intermediate magnetization plateaus at $1/5$ and $3/5$ of the saturation magnetization. Both plateaus remain discernible in the magnetization curve even at the slightly higher temperature $k_{\rm B}T/J = 0.05$, but rising thermal excitations eventually smear out all signatures of both intermediate plateaus at sufficiently high temperatures such as $k_{\rm B}T/J = 0.2$. Since the onset and breakdown of the intermediate $1/5$ and $3/5$ plateaus are associated with four field-driven QPTs, one might expect four associated finite peaks in the magnetic susceptibility at finite temperatures.  Surprisingly, none of these four peaks is clearly resolved in Fig.~\ref{fig5}(b) most likely because both plateau phases are characterized by relatively small excitation gaps. Instead, the intermediate 1/5 and 3/5 plateaus are identified by the pronounced suppression of the magnetic susceptibility, which nearly vanishes at the lowest temperature $k_{\rm B}T/J = 0.01$ within the narrow field intervals centered around $h/J \approx 0.85$ and $h/J \approx 1.35$, respectively.
Moreover, the magnetic susceptibility exhibits two distinct peaks at the lowest temperature $k_{\rm B}T/J = 0.01$. The round peak near $h/J \approx 1.1$ reflects the enhanced magnetic response of the spin-canted phase embedded in between two gapped plateau phases, while the sharp peak near $h/J \approx 1.82$ can be interpreted as a finite-temperature remnant of the field-driven QPT from the gapless spin-canted phase to the fully saturated state. The distinct physical origin  of these peaks is also reflected in their markedly different response with respect to the temperature. The broader peak near $h/J \approx 1.1$ remains well developed even at slightly higher temperatures such as $k_{\rm B}T/J = 0.05$ and $0.1$, whereas the sharp peak near $h/J \approx 1.82$ is rapidly suppressed and shifts toward lower magnetic fields.

\begin{figure}
\centering
\includegraphics[scale=0.28]{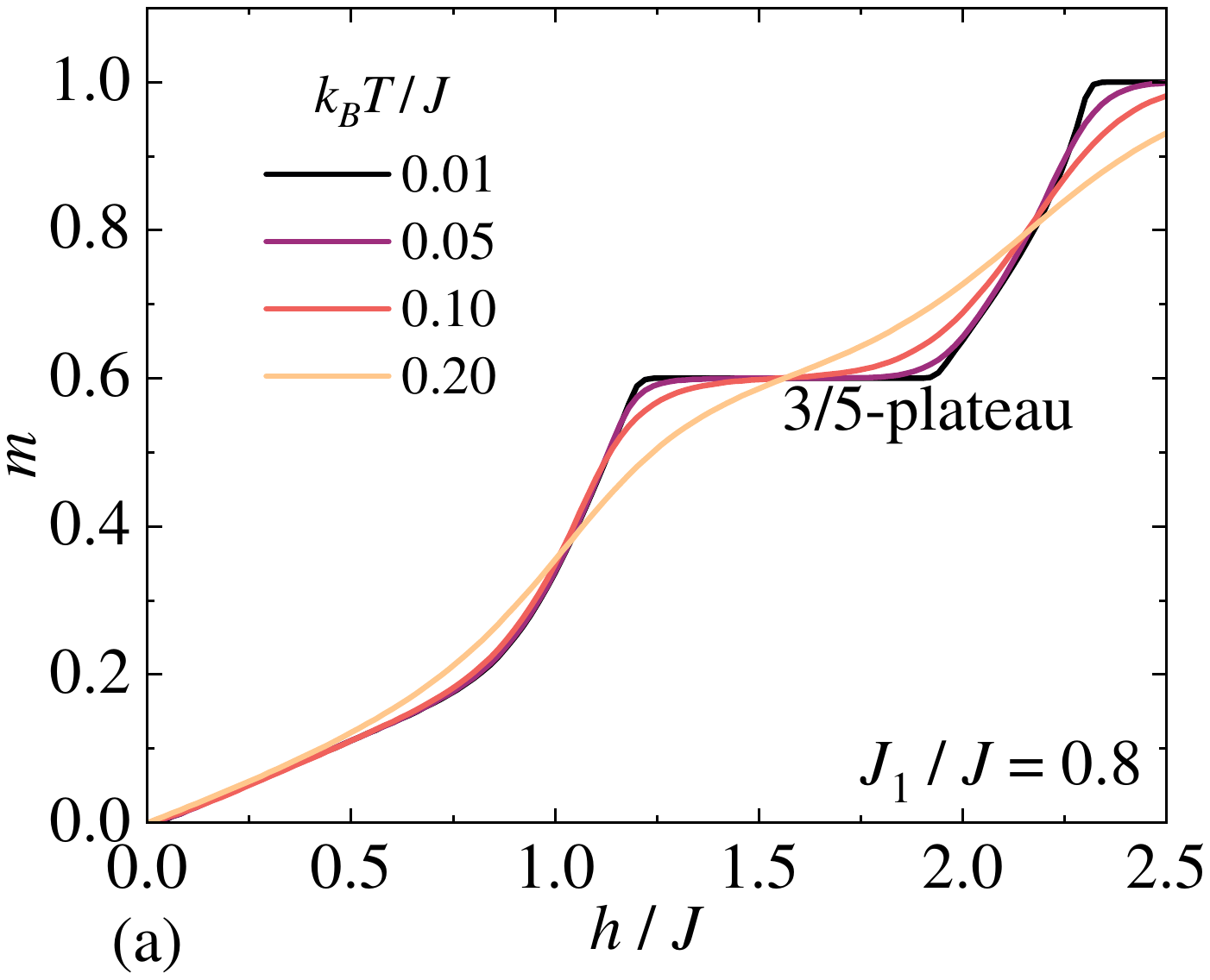}\\[-0.5cm]
\includegraphics[scale=0.28]{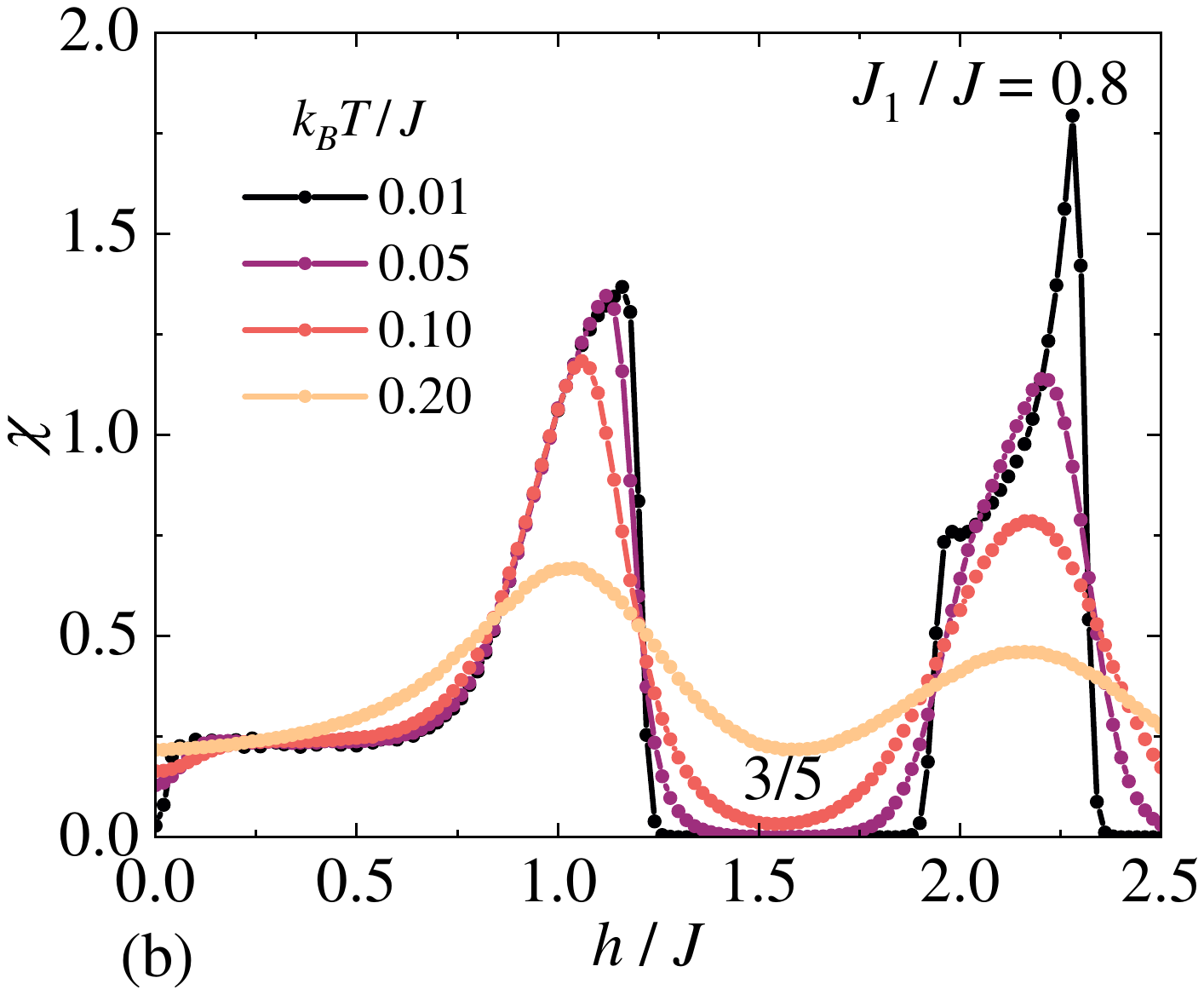}
\caption{Field dependencies of the magnetization normalized with respect to its saturation value (a) and the corresponding magnetic susceptibility (b) for the spin-1/2 Heisenberg model on the extended Lieb lattice at four distinct temperatures for $J_1/J=0.8$ as obtained from QMC simulations for the linear system size $L=8$ (320 spins).}
\label{fig6}
\end{figure}

Finally, we analyze in detail the case of $J_1/J = 0.8$, for which the field dependencies of the magnetization and magnetic susceptibility are plotted in Fig.~\ref{fig6} at four different temperatures. The isothermal magnetization curve presented in Fig.~\ref{fig6}(a) at the lowest temperature $k_{\rm B}T/J = 0.01$ clearly exhibits a sizable intermediate $3/5$ plateau extending over the field interval from $h/J \approx 1.2$ to $h/J \approx 1.9$, whereas the $1/5$ plateau is fully suppressed by the intervening gapless spin-canted phase. Owing to its relatively large excitation gap, the $3/5$-plateau phase clearly manifests itself up to moderate temperatures $k_{\rm B}T/J = 0.1$. A more detailed examination of the magnetic susceptibility shown in Fig.~\ref{fig6}(b) reveals three finite peaks at the lowest considered temperature $k_{\rm B}T/J = 0.01$. The first pronounced peak near $h/J \approx 1.2$ and the second peak with the shape of round shoulder centered around $h/J \approx 1.9$ can be interpreted as residual signatures of the field-induced QPTs associated with the onset and breakdown of the gapped $3/5$-plateau phase. The third peak located near $h/J \approx 2.28$ is remnant signature of the field-driven QPT from the gapless spin-canted phase to the fully saturated phase.  As the temperature increases, the latter two susceptibility peaks merge into a single broad maximum, which progressively shifts toward the center of the field interval corresponding to the spin-canted phase.

\section{Concluding remarks}\label{results}

We have investigated the magnetic properties of the spin-$1/2$ Heisenberg antiferromagnet on the extended Lieb lattice in the presence of an external magnetic field. Using the DMRG method, we calculated zero-temperature magnetization curves and established the ground-state phase diagram. Five distinct ground states were identified: three gapped phases corresponding in the magnetization curves to the intermediate $0$, $1/5$, and $3/5$ plateaus, a gapless spin-canted phase occupying the field regions between the gapped phases, and a fully saturated phase stabilized at sufficiently high magnetic fields. The onset and breakdown of the intermediate 1/5 and 3/5 magnetization plateaus are accompanied with the field-induced QPTs. The finite-temperature magnetic properties of the spin-$1/2$ Heisenberg antiferromagnet on the extended Lieb lattice were subsequently explored by means of QMC simulations, which revealed clear finite-temperature signatures of the field-driven QPTs in the magnetization and magnetic susceptibility. 
Although the QMC simulations provide no evidence for either continuous or discontinuous phase transitions at finite temperatures, the possibility of detecting a topological phase transition of Berezinskii-Kosterlitz-Thouless type at finite temperatures cannot be ruled out. To confirm the presence or absence of a topological phase transition would require a more thorough numerical analysis, which is however beyond the scope of this work. 


\printcredits

\section*{Declaration of competing interest}
The authors declare that they have no known competing financial interests or personal relationships that could have appeared to influence the work reported in this paper.

\section*{Acknowledgment}

This work was financially supported under the grant No. VEGA 1/0298/25, APVV-24-0091, and VVGS-2025-3497.

\section*{Data availability}

Data will be made available on request.



\begin{thebibliography}{10}
\bibitem{lieb86} \href{https://doi.org/10.1007/3-540-16473-1}{R. Liebmann, Statistical Mechanics of Periodic Frustrated Ising Systems, Springer, Berlin, Heidelberg, 1986.}
\bibitem{die04} \href{https://doi.org/10.1142/5697}{H. T. Diep, Frustrated Spin Systems, World Scientific, Singapore, 2004.}
\bibitem{kal11} \href{https://doi.org/10.1103/PhysRevB.84.174407}{A. Kalz, A. Honecker, and M. Moliner, Analysis of the phase transition for the Ising model on the frustrated square lattice, Phys. Rev. B 84 (2011) 174407.}
\bibitem{jin12} \href{https://doi.org/10.1103/PhysRevLett.108.045702}{S. Jin, A. Sen, and A. W. Sandvik, Ashkin-Teller criticality and pseudo-first-order behavior in a frustrated Ising model on the square lattice, Phys. Rev. Lett. 108 (2012) 045702.}
\bibitem{cac23} \href{https://doi.org/10.1103/PhysRevB.107.115143}{N. Caci, K. Kar\v{l}ov\'{a}, T. Verkholyak, J. Stre\v{c}ka, S. Wessel, and A. Honecker, Phases of the spin-$\frac{1}{2}$ Heisenberg antiferromagnet on the diamond-decorated square lattice in a magnetic field, Phys. Rev. B 107 (2023) 115143.}
\bibitem{sen04} \href{https://doi.org/10.1126/science.1091806}{T. Senthil, A. Vishwanath, L. Balents, S. Sachdev, and M. P. A. Fisher, Deconfined Quantum Critical Points, Science 303 (2004) 1490.}
\bibitem{lac10} \href{https://doi.org/10.1007/978-3-642-10589-0}{C. Lacroix, P. Mendels, and F. Mila, Introduction to frustrated magnetism, Springer, Berlin, Heidelberg, 2010.}
\bibitem{fan24} \href{https://doi.org/10.1038/s41535-024-00636-4}{Y. Fan, N. Xi, C. Liu, B. Normand, R. Yu, Field-controlled multicritical behavior and emergent universality in fully frustrated quantum magnets, npj QuantumMater. 9 (2024) 25.}
\bibitem{siv26} \href{https://doi.org/10.48550/arXiv.2606.10821}{D. Siv\'{y} and J. Stre\v{c}ka, Continuous and discontinuous transitions in the Ising-Heisenberg model on the extended Lieb lattice in a magnetic field, arXiv:2606.10821 (2026).}
\bibitem{fis59} \href{https://doi.org/10.1103/PhysRev.113.969}{M. E. Fisher, Transformations of Ising Models, Phys. Rev. 113 (1959) 969.}
\bibitem{bau11} \href{https://doi.org/10.1088/1742-5468/2011/05/P05001}{B. Bauer, L. D. Carr, H. G. Evertz, A. Feiguin, J. Freire, S. Fuchs, L. Gamper, J. Gukelberger, E. Gull, S. Guertler et al., J. Stat. Mech.: Theor. Exp. (2011) P05001.}
\end{thebibliography}

\end{document}